\documentclass[usenatbib]{basi}
\usepackage[T1]{fontenc}
\usepackage[british]{babel}
\usepackage[varg]{txfonts}
\usepackage{rotating}
\usepackage{dcolumn}

\begin{document}

\title[Study of ram pressure on NGC~2805]{Study of ram pressure effects on 
NGC~2805 in Holmberg~124}
\author[Mishra et. al.]{Alka Mishra,$^{1}$\thanks{E-mail:
alkam7@gmail.com}
N. G. Kantharia$^{2}$\footnotemark[1]\thanks{E-mail:ngk@ncra.tifr.res.in} and
D. C. Srivastava $^1$\footnotemark[1]\thanks{E-mail:dcs.gkp@gmail.com}\\
$^{1}$D.D.U. Gorakhpur University, Gorakhpur, India\\
$^{2}$National Centre for Radio Astrophysics, TIFR, Post Bag 3, Ganeshkhind, Pune, India}

\pubyear{2012}
\volume{40}
\pagerange{\pageref{firstpage}--\pageref{lastpage}}

\date{Received 2012 August 20; accepted 2012 December 13}

\maketitle

\label{firstpage}

\begin{abstract}
In this paper we present new H{\sc i} 21cm spectral line images of the poor
group of late type galaxies, Holmberg~124 made using archival data from 
the Giant Metrewave Radio Telescope (GMRT).  Holmberg~124 is a group of 
four late type galaxies: NGC~2805, NGC~2814, NGC~2820 and Mrk~108.  We 
detect spectral line emission from all the four galaxies and note several 
signatures of tidal interactions among the member galaxies.  Our results 
for the triplet (namely NGC~2820, NGC~2814 and Mrk~108) confirm the earlier 
results of Kantharia et al. (2005) notably the detection of a possible tidal dwarf 
galaxy to the north-east of NGC~2820.  Further, in these  images where 
the pointing center of the observations was changed, we have recovered 
most of the H{\sc i} emission in NGC~2805 as compared to Kantharia et al. (2005).  
We also report possible detection of small discrete clouds between NGC~2820 
and NGC~2805 which might be stripped H{\sc i} in the intragroup 
medium (IGrM). However, these need confirmation.  The H{\sc i} distribution of NGC~2805 
is asymmetric with peak H{\sc i} column densities seen along the southern spiral
arm and along a northern arc. Diffuse H{\sc i} is seen from the entire optical galaxy and
extends much further in the southern parts, especially in the south-west. 
An abrupt fall in H{\sc i} column densities is observed to the 
south of NGC~2820 which is a high inclination galaxy and to the north of 
NGC~2805 which is a low inclination galaxy.
Vigorous star formation has been observed along the southern spiral arm of NGC~2805. 
Based on these new H{\sc i} images, we support the scenario given by Kantharia et al. (2005) that
both tidal interactions and ram pressure are currently playing a role in the 
evolution of the triplet galaxies.  From the observed northern H{\sc i} arc and 
extensive star formation
in the southern spiral arm of NGC~2805 and additionally the systemic velocities 
of the four galaxies, we suggest ram pressure effects are also playing
a role in the evolution of NGC~2805.  We believe that the 
H{\sc i} in the north  of NGC~2805 is being 
compressed as it moves in the IGrM and that it is moving along a direction
in the north-east close to the line-of-sight so that the entire disk
is encountering the IGrM.  This interaction with the IGrM 
would have triggered star formation in the southern spiral arm.  
This model for NGC~2805 succeeds in explaining 
the compressed H{\sc i} in the north,  widespread star formation and the 
diffuse H{\sc i} detected in the south/south-west of the optical galaxy.
Deep X-ray observations of the IGrM and deeper H{\sc i} observations
sensitive to larger angular scales will be very useful in furthering
the understanding of this interesting group of galaxies.  
\end{abstract}


\section{Introduction}\label{s:intro}

Environment plays an important role in the evolution of galaxies. 
Tidal interactions \citep{toomre72}, galaxy harassment \citep{moore96}, 
removal of gas via ram pressure stripping \citep{gunn72}, viscous stripping
\citep{nulsen82} and major \& minor mergers \citep{toomre77}
play a key role in galaxy evolution and modify their  characteristics, like morphology, 
kinematics  and star formation rates(SFR). While the importance of the various physical 
mechanisms in determining the evolution of the member galaxies is well understood in 
cluster environments; we are still understanding the importance of the various processes 
in groups of galaxies.  For example, in cluster environments, it is known 
that ram pressure stripping is the main mechanism for stripping the galaxies of
their gas near the cluster center and that harassment plays an important role in 
stripping the galaxies of gas mass near the outskirts and intermediate distances from 
the cluster center.  However, while tidal interactions are believed to play a major 
role in group environs due to the lower relative velocities and lower IGrM 
densities - it is not clear if this is the only major process active in group environs.  

One major reason for the need to look for additional physical processes which
are active in group environs is the presence of morphological signatures 
which have traditionally been attributed to ram pressure stripping if the galaxy 
was located in cluster environs. For example, in the galaxies which are found 
embedded inside cluster environs an abrupt drop in H{\sc i} column density giving rise 
to a smooth edge or H{\sc i} depletion are believed to be caused by their motion in the
medium.  However, when tidal interaction is invoked to explain similar
signatures observed in group members, there are difficulties in
explaining all the observed features. However since such observed cases are 
small in number, we still have much to learn about these systems  (\citealt{verdes01};
 \citealt{kantharia05}; \citealt{chung07}; \citealt{sengupta07};  
\citealt{rasmussen12}). 
Several simulations of such systems for a range of physical parameters 
have shown ram pressure to lead to observable signatures on the galaxies 
(\citealt{vollmer01}; \citealt{roediger06}). A study of this aspect for the poor 
group of late type galaxies, Holmberg~124, has been carried out by Kantharia 
et al. (2005) and in the present investigation we extend it further.
 
Holmberg~124 consists of four late type galaxies - NGC~2820, NGC~2805, 
NGC~2814 and Mrk~108 and has been subject of extensive study  to name a few;  radio
continuum observations (\citealt{van der hulst85}; \citealt{kantharia05}) H{\sc i} observations 
(\citealt{reakes79}; \citealt{bosma80}; \citealt{kantharia05}) and  photometric observations
(\citealt{bosma80}; \citealt{artamonov94}).  All the galaxies show several 
morphological and/ or other signatures attributable to tidal interactions.  
Besides, they show additional signatures which demand an alternative 
explanation such as ram pressure stripping \citep{kantharia05}.   
While the extensive H{\sc i} study of the group by Kantharia et al. 
(2005) using GMRT clearly detected several tidal and ram pressure events in the galaxies  
NGC~2820, NGC~2814 and Mrk~108 which are referred to as triplet in rest of the paper; 
owing to NGC~2805  being located close to the half power point in the GMRT primary beam 
during the observations; the reduced sensitivity did not recover all the H{\sc i} in the galaxy.  
The current data downloaded from the GMRT  archives, is free from this constraint with the 
pointing centre being between NGC~2820 and NGC~2805. 

NGC~2805 is an Sd galaxy seen nearly face-on and is the brightest member in the group
Holmberg~124. The H{\sc ii} regions in NGC~2805 appear to be distorted 
on the side of the galaxy opposite to the companion \citep{hodge75}.  H{\sc i} has been detected 
from the galaxy (\citealt{reakes79}; \citealt{bosma80}; \citealt{kantharia05}) 
and Bosma et al. (1980) report that 
the outer H{\sc i} layers are warped.  A spectroscopic study of a few galaxies including NGC~2805 
have been described by Ganda et al. (2006). This investigation reveals that the galaxy seems to 
be also optically disturbed, since the spiral arms appear to be broken up into
segments. They have employed SAURON spectrograph  and have studied the
central parts of the galaxy and detect slow projected stellar
velocities and a central drop in velocity dispersion. They conclude
that this could be due to the lack of a classical stellar bulge and
the presence of small-scale structures (nuclear star clusters, inner
rings, inner bars). The gas has a clumpy distribution and rotates
consistently with the stars. Low values of the line ratio [O{\sc iii}]/H$\beta$ 
are observed over the central 33 $\times$ 41
arcsec$^{2}$ of the galaxy, possibly indicating ongoing star formation
\citep{ganda06}.  Boker et al. (2002) have reported that the galaxy hosts a compact
luminous stellar cluster in the centre and  performing an analytical
fit to the surface brightness profiles have obtained a size for the
nuclear star cluster of 8.2 pc and an absolute I-band magnitude of
-13.32 for a distance of 28.1 Mpc to NGC~2805.
The H$\alpha$ rotation curve studies of NGC~2805 by Garrido et al.
(2004) reveal that it is quite regular and symmetric and reaches 70 km
s$^{-1}$ at 30 arcsec, then rises slowly up to 85 km s$^{-1}$ (beyond 120 arcsec).
They find that on the isovelocities map, the position of the major axis is changing with
radius and the approaching side rises beyond 120 arcsec up to 120 km
s$^{-1}$; confirming the warps pointed out by the Bosma et al. (1980).
The equipartition magnetic field in NGC~2805 has been estimated to be
1.86 $\mu$G  \citep{fitt93}.

In this paper, we present new H{\sc i} images of the group Holmberg~124 especially
the almost face-on spiral member, NGC~2805.  We also present the kinematics
of NGC~2805 and discuss the observed morphology.  We combine our new sensitive 
radio continuum maps at 20cm 
of the triplet with the 325 MHz maps from Kantharia et al. (2005) and
present the spectral index distribution across the triplet members.
We compare the radio emission with other wavebands and comment on the 
evolution of the group especially the physical processes acting on
NGC~2805. Based on our new  images and using data from literature, we refine 
the model given by Kantharia et al. (2005) for the poor group of galaxies
Holmberg~124. We assume  the distance of the group to be 25 Mpc with the
estimated velocity dispersion along the line of sight as  162$\pm$73
km s$^{-1}$  \citep{kantharia05}.

\begin{table}
\begin{center}
\caption{Galaxy parameters}
\begin{tabular}{@{}lllll@{}}\hline

Properties       &   NGC~2805    &NGC~2820 &  NGC~2814 & IC~2458 \\
                 &              &         &           & (Mrk~108)\\ \hline 
Optical size$^{a}$&  $6.3'\times4.8'$ & $4.3'\times0.53'$ & $1.2'\times0.3'$ & $0.5'\times0.2'$\\
Classification$^{a}$&  SAB(rs)d & SB(s)c &    Sb     & I0 pec\\
Radial velocity$^{a}$
(km s$^{-1}$)       & 1725      &1574    &    1592   & 1534\\
Optical linear size (kpc) $^{b}$&45 &31.2&     8.7  &  3.6 \\ 
B band magnitude
(absolute)$^{c}$   & $-$20.84     &$-$20.70  &   $-$18.88  & $-$16.89\\ 

log FIR (W~m$^{-2}$)$^{d}$ $\times$ 10$^{-14}$ &10.92$\pm$1.3 & 23.30$\pm$1.5 & 10$\pm$0.4 & ----   \\
Single dish H{\sc i} flux density  &              &         &           &     \\  
(Jy~km s$^{-1}$)$^{e}$& 90.07$\pm$0.68&   ----        &  ---- &  ----   \\
\hline
\end{tabular}\\[5pt]
\begin{minipage}{10 cm}

\small Notes: (a) NASA/IPAC Extragalactic Data base (NED)\\
              (b) $\theta$ $\times$ distance, where $\theta$ is the major axis of the galaxies in arcmin.\\
              (c) Hyperleda \\
              (d) log FIR ($W~m^{-2}$) = $1.26\times10^{-14}[2.58S_{60 \mu m} + S_{100 \mu m}/Jy]$\\
                  The flux density values for 60 and 100 $\mu$m have been taken from NED. 
                  For Mrk~108, 60 and 
                  100 $\mu$m flux density values are not available.\\
              (e) \citet{lang03}; single dish flux densities are not available for the other galaxies. \\
\end{minipage}
\end{center}
\end{table}

\begin{figure}
\centerline{\includegraphics[height=9.0cm,width=9.0cm]{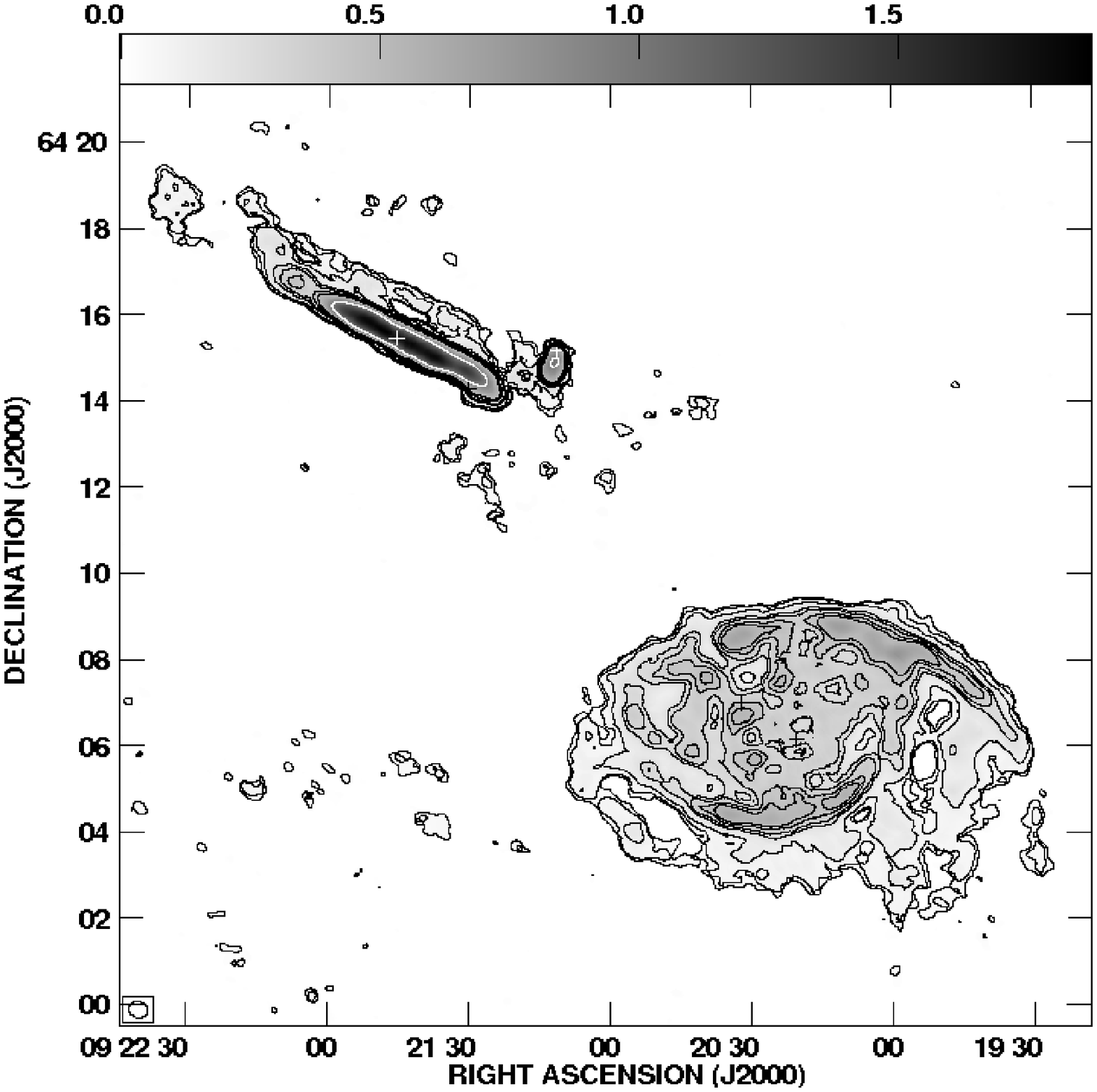}}
\centerline{\includegraphics[height=9.0cm,width=9.0cm]{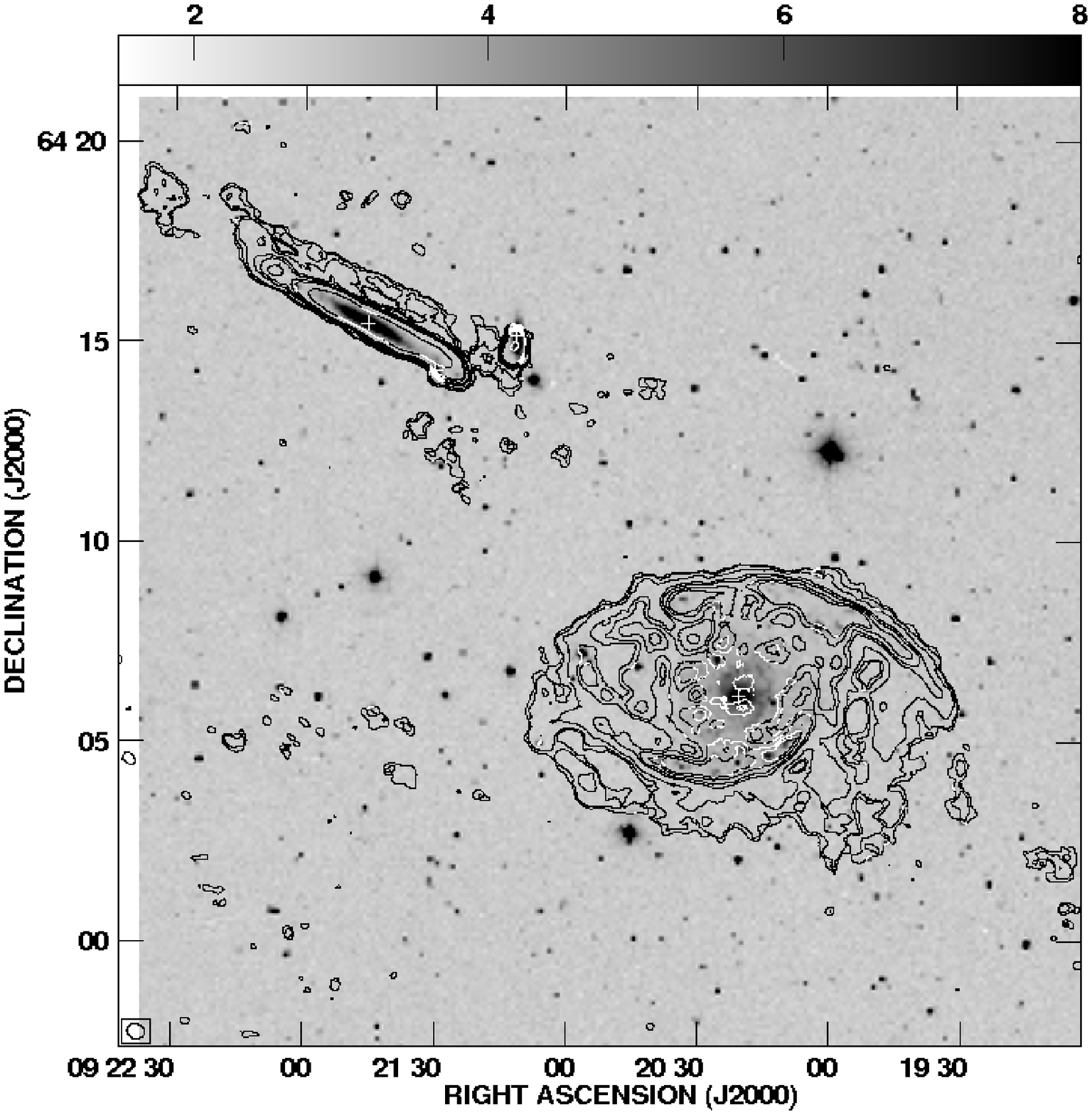}}
\caption{{\bf(a)}(top panel)H{\sc i} column density contours  and  grey-scale at an angular
resolution of $26'' \times 22''$. Crosses mark the optical positions of the
four group members. Note the large structure to the north-east of NGC~2820 which
we believe is a tidal dwarf galaxy. {\bf(b)}(lower panel) The H{\sc i} column density contours overlaid 
on the DSS B band (grey scale) image at an angular resolution of $26'' \times 22''$. 
The contours are plotted at 0.36$\times$(3, 6, 12, 24, 40, 48, 96, 192, 220)$\times$10$^{19}$ 
atoms cm$^{-2}$. }
\label{fig:Holm124group}
\end{figure}

\section{Data analysis}\label{s:obser}
\footnotetext[1]{AIPS is distributed by the National Radio Astronomy Observatory, which
is a facility of the National Science Foundation operated under cooperative
agreement by Associated Universities, Inc.}

The data analyzed and presented here are obtained from the GMRT archives. 
GMRT is an array  of 30 antennas, each of 45 meter in diameter spread over 
a maximum baseline length of 25 km \citep{swarup91}. 
An observing bandwidth of 4 MHz centered at 1411.2 MHz (which 
corresponds to a heliocentric velocity 1725 km~s$^{-1}$) was used. The band 
was divided into 128 spectral channels giving a channel spacing of 6.63 
km~s$^{-1}$.  Flux calibration was done using scans on the standard calibrator 3C147 
and 3C286, and  which were observed at the start and end of the observing run. 
Phase calibration was done using the calibrator source 0834+555 
($\sim$ $9^{\circ}$ away from the target source) which has $S_{1.4}$ GHz = 8.3 Jy. 
Phase calibrator was observed once in every 30 minutes.  Bandpass calibration 
was done using 3C147.  The total on-source time was $\sim$ 8 hours. The data was 
reduced using standard tasks in NRAO AIPS$^1$. Bad data due to dead antennas or 
radio frequency interference (RFI) and antennas with significantly lower gain than 
others were flagged. The calibrated data were then used to make the 21cm radio 
continuum images by averaging the line-free channels.  This data was imaged using IMAGR.
The final continuum image with a resolution of  $6''\times 4''$, PA $= 87.8^{\circ}$.
was made using uniform weighting and had an rms noise of 0.067 mJy/beam.  For making the
H{\sc i} line emission images, the calibrated  data were continuum subtracted using the 
AIPS tasks UVSUB. The task IMAGR was then used to get the final three-dimensional 
deconvolved H{\sc i} data cubes. From these cubes, the moment maps giving H{\sc i} column density, 
the intensity-weighted H{\sc i} velocity field and the intensity-weighted dispersion 
were extracted using the AIPS task MOMNT. 
Data cubes were made at various cutoffs in the uv plane such as 0-20 
$k\lambda$, 0-15 $k\lambda$, 0-10 $k\lambda$ and 0-5 $k\lambda$ by using 
natural weighting. The corresponding angular resolutions are $13'' \times 11'', 17'' \times 15'', 
26'' \times 22'', 41'' \times 38''$.  RMS noise per channel for the 
different maps was 0.68 mJy/beam, 0.79 mJy/beam, 0.92 mJy/beam and  1 mJy/beam respectively. 
While the high resolution images fragmented the large scale emission; the lowest
resolution image only smoothened the detected features.  Thus, the image made
with a resolution of $26''\times 22''$, PA $= 70.3^{\circ}$ was used for 
subsequent analysis since it was sensitive to all the detected H{\sc i} and also 
retained sufficient resolution to resolve the features within the large disk.  
Peak H{\sc i} column density (as derived from the $26''\times 22''$, PA $= 70.3^{\circ}
$ resolution image) in NGC~2805 is about 10$^{20}$ atoms cm$^{-2}$. 
We use the image with an angular resolution of $26''\times 22''$, 
PA $= 70.3^{\circ}$ in rest of the discussion unless specified otherwise.  

\begin{table}
\begin{center}
\caption{RMS noise in the images at different resolutions.}
\begin{tabular}{@{}lccccc@{}}\hline
  No         &  UV cutoff  & Beam size & Position angle  & Continuum  & Spectral channel \\  
             &   k$\lambda$ & arcsec$^{2}$  & degree &  mJy/beam &  mJy/beam \\
\hline
1 &  Full-range  & 6$\times$4  &  87.8 & 0.067  & -  \\
2 &  0-20       & 13$\times$11 &  69.7 & 0.079  & 0.68   \\
3 &  0-15       & 17$\times$15 &  79.1 & 0.100  & 0.79   \\
4 &  0-10       & 26$\times$22 &  70.3 & 0.130  & 0.92  \\
5 &  0-5        & 41$\times$38 &$-$50.0& 0.230  & 1.00  \\ \hline
\end{tabular}
\end{center}
\end{table}

\section{Results and discussion}\label{s:resul}
\subsection {H{\sc i} emission}
21cm H{\sc i} emission is detected from all the group members (NGC~2805, NGC~2820,
NGC~2814 and Mrk~108). Fig. \ref{fig:Holm124group}(a) shows the column density 
of H{\sc i} in the member galaxies. In Fig. \ref{fig:Holm124group}(b), the column
density of H{\sc i} is shown superposed on the DSS optical image. All the members of 
the group are H{\sc i} rich and the H{\sc i} disk is more extended than 
the optical galaxy. H{\sc i} emission is also detected from several other features such 
as the bridge.  
In the case of NGC~2820, the H{\sc i} morphology is asymmetric 
with a large H{\sc i} loop seen to the north-west of NGC~2820 and a bridge is detected 
connecting NGC~2820 and Mrk~108 with NGC~2814.  This is the first detection to 
the best of our knowledge of the entire H{\sc i} bridge connecting the triplet members.  
However, detection of part of the bridge has been reported earlier
by Kantharia et al. (2005). 
A streamer positionally emerging from NGC~2814 but kinematically distinct from it
is also detected here.  The H{\sc i} disk in NGC~2814 is displaced to the south relative 
to the optical disk. A H{\sc i} tail is seen to the west of NGC~2814. We confirm the 
detection of the tidal dwarf galaxy located to the north-east of NGC~2820 that 
Kantharia et al. (2005) had first reported.  All features that were reported by 
Kantharia et al. (2005) are visible in our map - additionally we detect the entire 
H{\sc i} bridge and small H{\sc i} clouds around the triplet  Fig. \ref{fig:Holm124group}(a)
and Fig. \ref{fig:Holm124group}(b). Since these are low column density clouds, 
they need to be confirmed.  We note that such H{\sc i} clouds residing in
the intragroup environs have been detected in several groups (\citealt{verdes01}; 
\citealt{dahlem05}; \citealt{kilborn06}).

\begin{figure}
\centerline{\includegraphics[height=10.0cm,width=10.0cm,angle=-90]{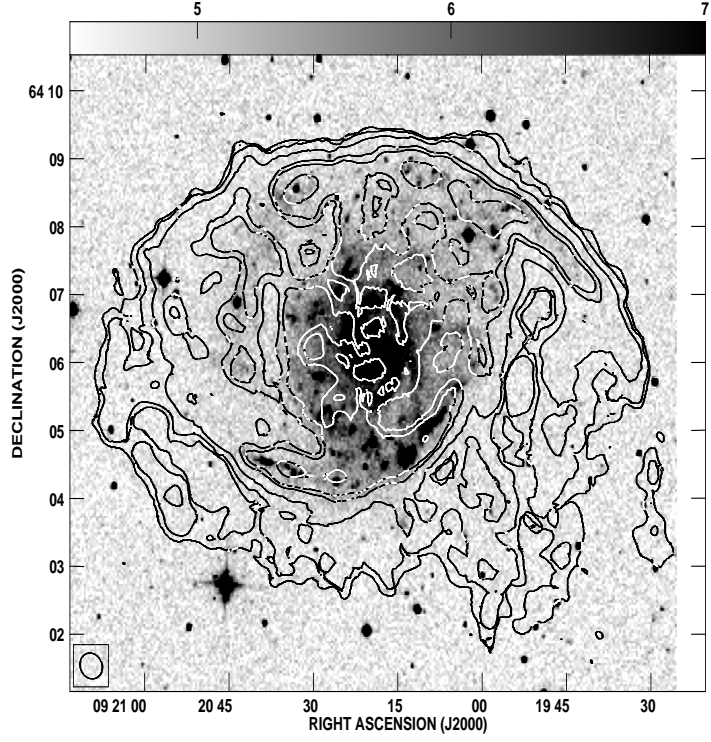}}
\caption{Zoomed-in column density image of NGC~2805   
at an angular resolution of $26''\times 22''$, PA $= 70.3^{\circ}$ overlaid 
on a DSS B band image. 
The contour levels are 0.36$\times$(3, 6, 12, 24, 30, 48)$\times$10$^{19}$ atoms cm$^{-2}$. 
Note the high column densities regions in the northern arc and the southern 
spiral arm in addition to the star forming ridge in the
southern spiral arm. }
\label{fig:ngc2805m0}
\end{figure}

The H{\sc i} disk of NGC~2805 upto a column density cutoff of 1.08$\times$10$^{19}$ cm$^{-2}$
is 8.3$^\prime$ $\times$ 6.6$^\prime$ which is $\sim$ 30 \% more extended 
than the optical disk.  The integrated line strength is  86.9$\pm$1.2 
Jy km s$^{-1}$ and the H{\sc i} mass estimated from this is 12.5 $\times$ 
10$^{9}$ M$_{\odot}$. The integrated H{\sc i} intensity map and H{\sc i} kinematics of 
NGC~2805 overlayed on the digitized sky survey (DSS) blue band image are shown in 
Figs. \ref{fig:ngc2805m0}, \ref{fig:ngc2805m1}.  As seen in Fig. \ref{fig:ngc2805m0},
enhanced H{\sc i} column densities are seen to be confined to the optical disk of 
the galaxy with the highest column densities along the northern star 
forming ridge and the southern spiral arm.  H{\sc i} extends well beyond the 
optical disk of the galaxy. Low column density gas is seen to extend beyond the southern 
ridge of star formation. This gas appears to be patchy and flocculent.  Interestingly, H{\sc i} 
does not extend beyond the northern ridge and there seems to be an abrupt fall in the column 
densities. The column densities over most of the disk is uniform.  In Fig.
\ref{fig:channelmap}, the spectral channel maps of the H{\sc i} from NGC~2805 are shown.  
The disk is rotating with the north-west receding from us and the south-east approaching 
us. The SDSS gri image shows an 
intense bow-like ridge to the south (also seen in other images) which ends abruptly in 
the south-west. The morphology is reminiscent of star formation being triggered by compression 
such as by ram pressure.  The H{\sc i} emission detected
beyond this ridge might not arise in the disk of the galaxy but as the galaxy moves in 
the IGrM, is being blown away by ram pressure. The observations 
suggests that the extended gas might be non-coplanar with the NGC~2805 disk. 

\subsection {H{\sc i} kinematics}
\begin{figure}
\centerline{\includegraphics[height=10.0cm,width=10.0cm,angle=-90]{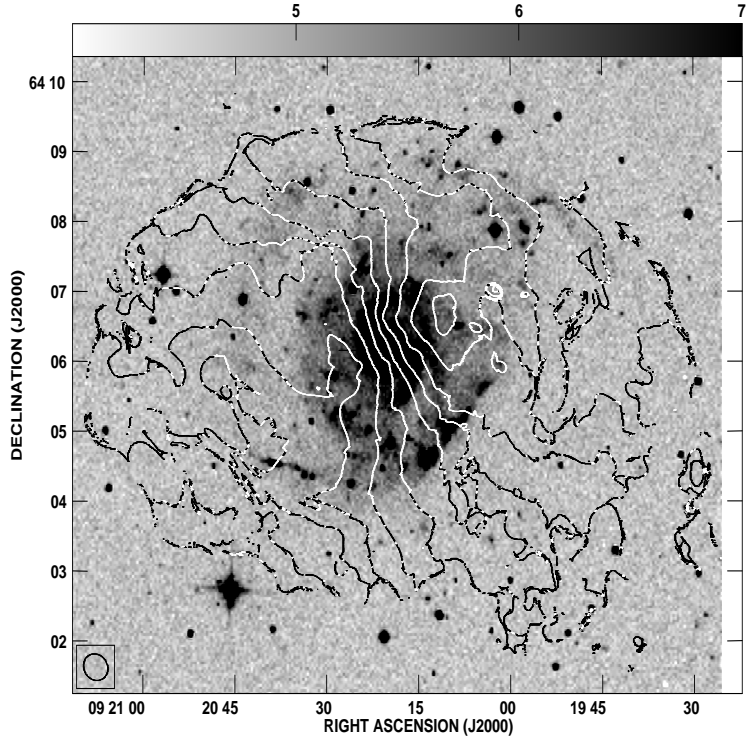}}
\caption{The H{\sc i} velocity field (first moment map) of NGC~2805 at an angular 
resolution of $26''\times 22''$, PA $= 70.3^{\circ}$ overlaid on a DSS B band image.
The contours are plotted in steps of 10 km s$^{-1}$  and range from 1680 km s$^{-1}$ 
to 1760 km s$^{-1}$ }
\label{fig:ngc2805m1}
\end{figure}

\begin{figure}
\centerline{\includegraphics[width= 14cm,angle=-90]{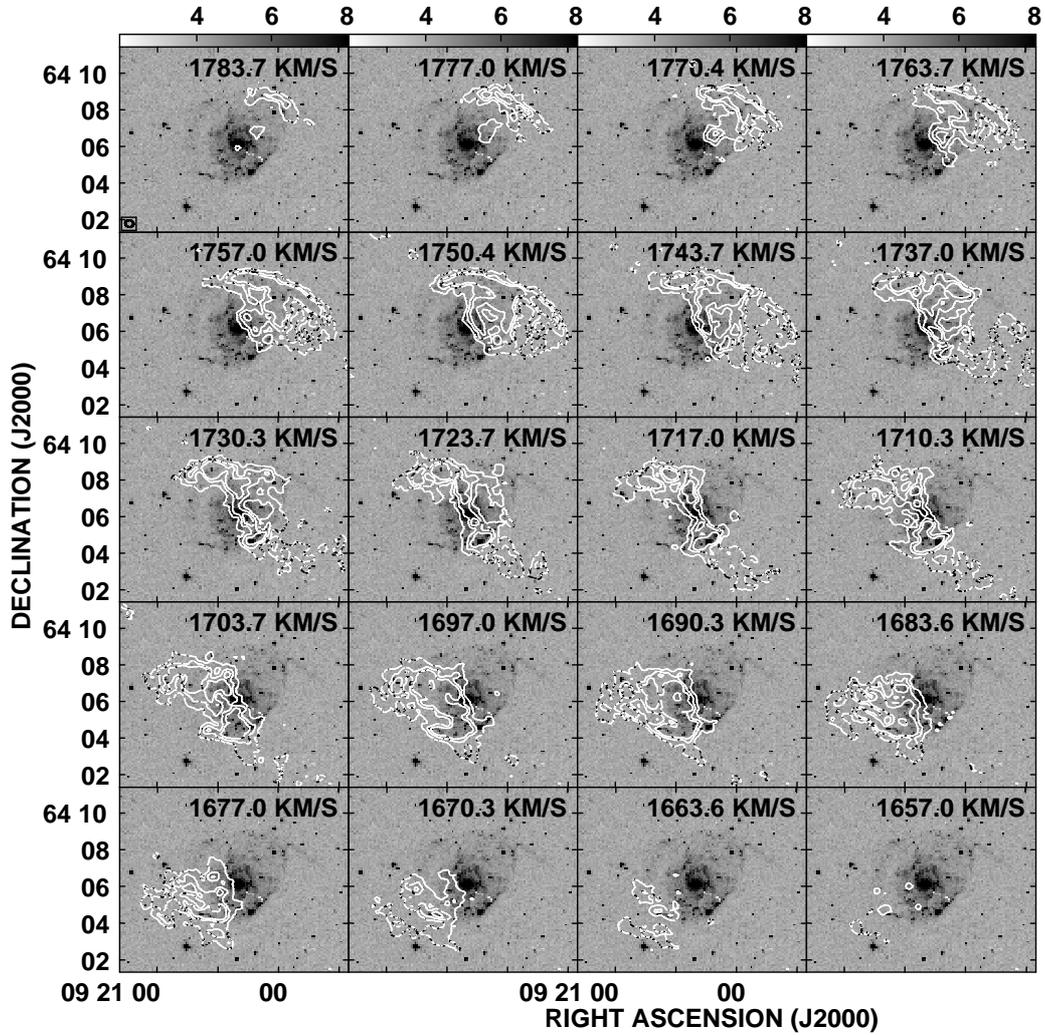}}
\caption{The H{\sc i} channel maps of NGC~2805 overlaid on the DSS B-band optical image, in grey scale.
The contour levels are 1 mJy beam$^{-1}$ $\times$ (3, 6, 9). The
heliocentric velocity  of the spectral channels are indicated in the upper left hand corner.}
\label{fig:channelmap}
\end{figure}

Fig. \ref{fig:ngc2805m1} shows the velocity field of NGC~2805.  The gradient in the 
velocity field from the north-west to the south-east is clearly seen. The H{\sc i} velocity 
field is fairly regular inside the optical disk but appears distorted beyond it; which 
is quite often a signature of a warp in the H{\sc i} disk.  
Several galaxies which show such features are well-fitted 
by a tilted ring model. Wiggles are also visible on smaller scales inside the optical disk. 
Fig. \ref{fig:channelmap} lets us better understand the connection between the
optical and H{\sc i} structures.  Notice the low column density gas in 
the west, south-west of the optical disk especially in the spectral channels with velocities 
ranging from 1744 km~s$^{-1}$ to 1710 km~s$^{-1}$.  This tenuous gas shows 
peculiar morphological and kinematic characteristics raising 
interesting possibilities on its origin.  Notice the cloud near the centre of galaxy
detected near velocities $1784-1770$ km~s$^{-1}$ in Fig. \ref{fig:channelmap} which
might be extraplanar.

\subsection {H{\sc i} Velocity dispersion}
\begin{figure}
\centerline{\includegraphics[height=9.0cm,width=9.0cm,angle=-90]{MOM2NGC2805MAP.eps}}
\centerline{\includegraphics[height=9.0cm,width=9.0cm,angle=-90]{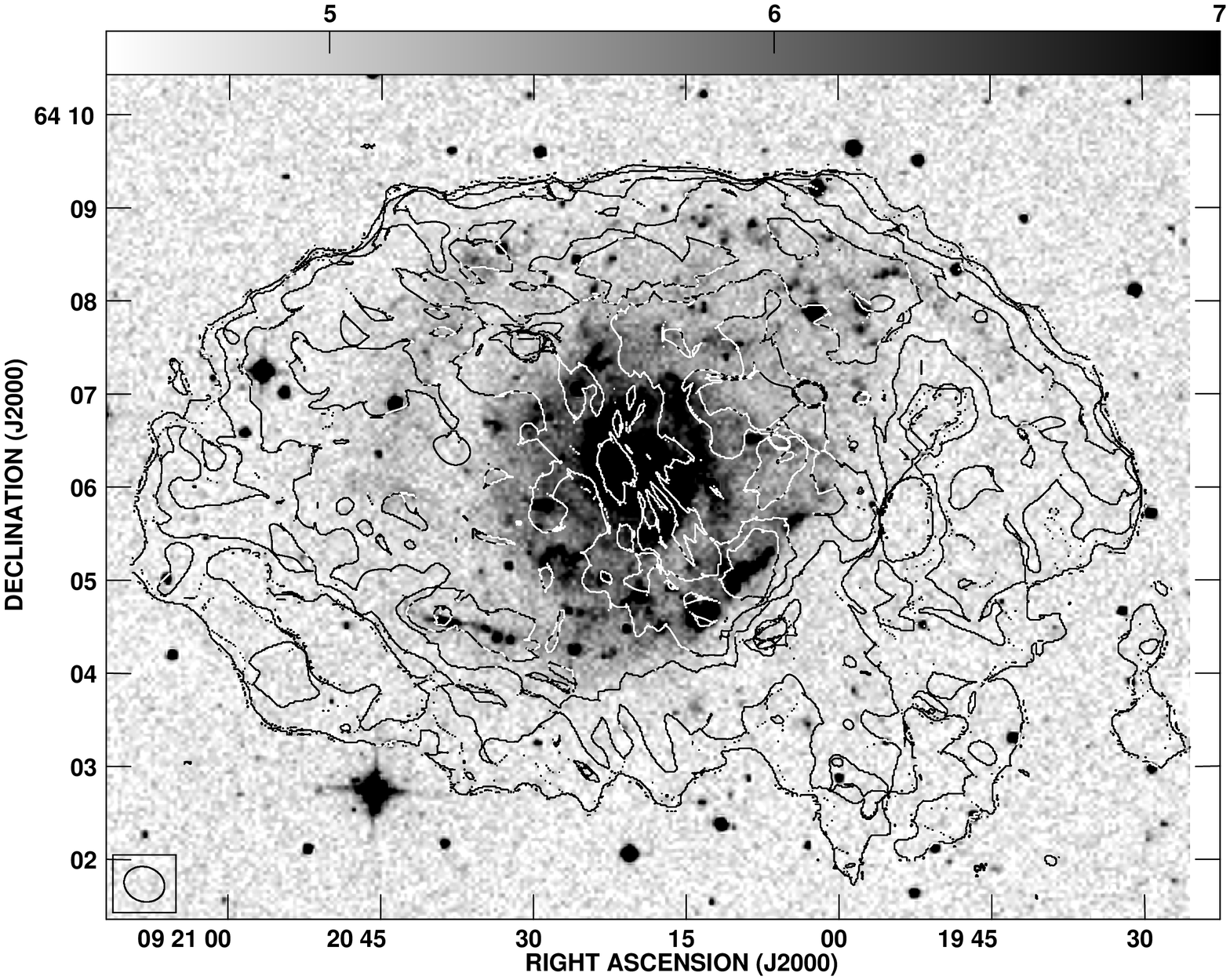}}
\caption{{\bf(a)}(top panel) The H{\sc i} intensity-weighted velocity dispersion 
contours  and  grey-scale at an angular resolution of $26''\times 22''$, 
PA $= 70.3^{\circ}$.{\bf(b)}(lower panel) The H{\sc i} intensity-weighted velocity 
dispersion contours overlaid on the DSS B band (grey scale). The contours are 
in steps of 2 km s$^{-1}$ and range from 2 to 20 km s$^{-1}$.}
\label{fig:ngc2805m2}
\end{figure}

Intensity-weighted velocity dispersion (second moment) map of NGC~2805 is shown in Fig. 
\ref{fig:ngc2805m2}. The dispersion ranges from 2 to 20 km~s$^{-1}$ with the largest 
dispersion seen in three distinct regions in the galaxy. The southern region showing 
a wide line is coincident with an intense star forming region in the southern arm. 
However the two other broad line regions are not coincident with intense star forming 
regions.  One of them lies to the south of the northern arc in NGC~2805
and the other is coincident with the star forming region in the northern arc which has 
large H{\sc i} column densities. The central broad line region is coincident with
the possible extraplanar cloud that was mentioned in the previous section. 
The central regions of the galaxy show lines of 
dispersion $\sim 15$ km~s$^{-1}$. Parts of the optical disk, the southern ridge and northern 
broken ring show dispersion of $\sim 12$ km~s$^{-1}$.  
Rest of the gas mass which is likely associated 
with the disk has a dispersion of about 9 km~s$^{-1}$.

The lines from  diffuse gas in the southern region outside the star
forming ridge are narrow and have a line width of 3 to 6 km/s.  This gas has no star
formation associated with it. Moreover as noted earlier, there is a 
possibility of this gas being extraplanar probably in
the process of being gently stripped due to ram pressure effects.

\begin{figure}
\centerline{\includegraphics[height=10.0cm,width=10.0cm,angle=-90]{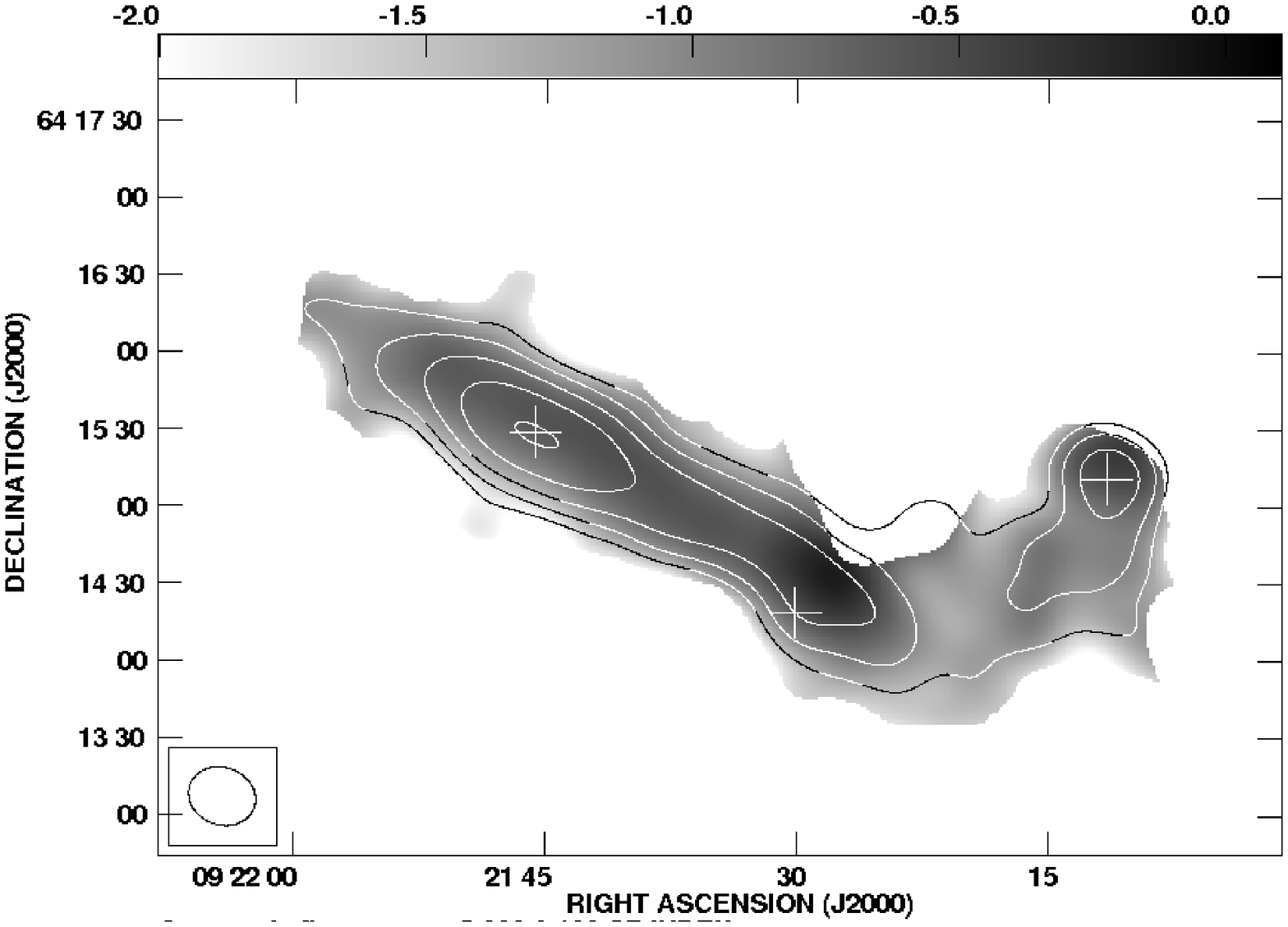}}
\caption{Spectral index ($\alpha_{325}^{1420}$) distribution (in grey scale) across
the triplet of galaxies superposed on the 1.4 GHz radio continuum contours at 
resolution of $6''\times 4''$, PA $= 87.8^{\circ}$. Counter 
levels are 0.3 $\times$ ($-$3, 3, 6, 12, 24, 48) mJy/b.}
\label{fig:ngc2820si}
\end{figure}

\subsection{Radio continuum}
We detect radio continuum emission at 21cm from the triplet. We also detect the 
bridge of radio emission connecting NGC~2820/Mrk~108 with NGC~2814, which was first detected 
by \citet{van der hulst85}.  But, no radio continuum emission was detected down to 
a  3$\sigma$ flux density limit of about 0.4 mJy/beam (for beam size of $26'' \times 22''$) from NGC~2805
at 21cm. Hence using the NVSS flux density of NGC~2805 at 21cm and the 90cm
flux density from  Kantharia et al. (2005), we estimate the spectral index
to be $-$0.8$\pm$0.2.  In order to study the spectral index distribution in the bridge and
the triplet of galaxies we use the 21cm image along with the 90cm image
from Kantharia et al. (2005)  and the results are shown in 
Fig. \ref{fig:ngc2820si}.  Kantharia et al. (2005) who used the total 
emission at 20cm and at 90cm estimated a  spectral index of $-1.8$ 
for the bridge emission.  This is similar to the distribution shown in 
Fig. \ref{fig:ngc2820si} which reveals that most of the bridge shows a spectral index steeper 
than $-1.5$.  While the spectral index close to Mrk~108 and the optical disk of NGC 
2814 seem to be about -0.6; the spectral index of most of the emission from NGC~2820 
is found to be steeper than Mrk~108.

\begin{figure}
\centerline{\includegraphics[height=10.0cm,width=10.0cm,angle=-90]{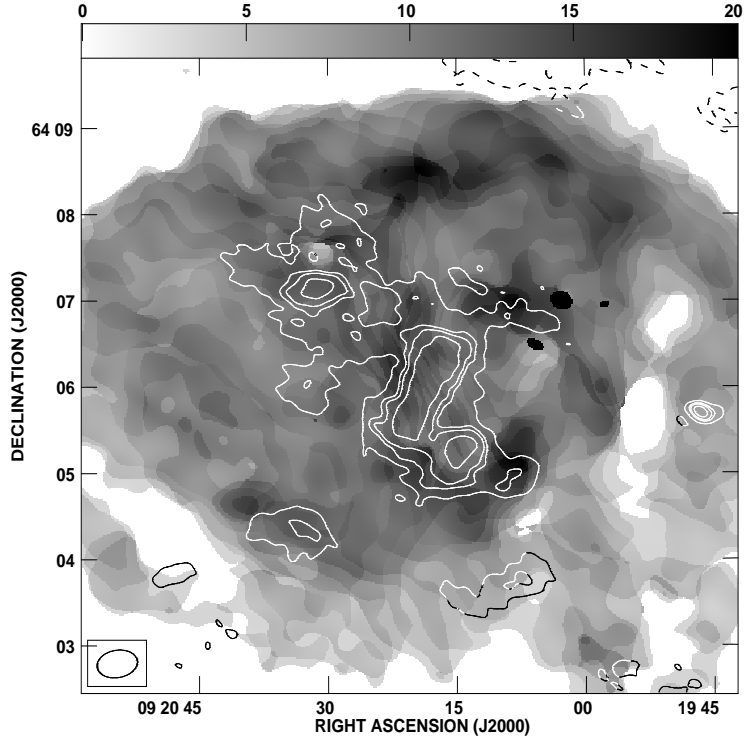}}
\caption{Radio continuum emission from NGC~2805 at 325 MHz
(shown in contours) superposed on the velocity dispersion image.  Counter levels
are 2.5 $\times$ ($-$2, $-$1.4, 1.4, 2.0, 2.8, 4.0) mJy/b and the grey scale ranges from
0$-$20 km/s.} 
\label{fig:ngc2805_325mhz}
\end{figure}

We present a comparison of 325 MHz radio continuum emission (from Kantharia et al. 2005) 
with the H{\sc i} emissions in Fig. \ref{fig:ngc2805_325mhz}. It turns out that 
the radio continuum emission at 325 MHz, which is confined to the optical disk, 
is seen to be coincident with the two of the broad line regions seen in H{\sc i}.  
Incidentally, the radio continuum emission is detected from the regions 
which show higher H{\sc i} velocity dispersion in the galaxy 
[Fig. \ref{fig:ngc2805_325mhz}]. However, no radio continuum emission at 
325 MHz is detected from the northern arc which shows 
weak star formation. 

Artamonov (1994) have pointed out the chain of blue condensation 
along the spiral arms in NGC~2805 and concluded that the median age of these blue objects
with dust correction is  10$^{6}$ -10$^{7}$ yr. This interpretation is 
supported by the morphologies seen in the NIR by 2MASS and in the UV by GALEX 
images of NGC~2805. Both these show excess nuclear emission but the southern 
spiral arm is barely visible in the NIR image.  In the UV image, young star 
forming regions are visible throughout the inner spiral arms and along the 
spiral bow-shaped ridge.  This suggests that star formation
has been recently triggered throughout this galaxy, probably due to an external 
interaction in the group environs.  The burst of star formation could have been triggered 
less than 10$^7$ years ago, thus explaining the absence of radio continuum emission being 
associated with all the regions.   Assuming that the entire emission
detected from NGC~2805 at 1.4 GHz by NVSS is non-thermal synchrotron 
emission due to star formation, we estimate a global
SFR of 0.86 M$\odot$yr$^{-1}$ for NGC~2805.
The SFR per unit area of the NGC~2805 disk is estimated to be $0.22\times10^{-8}$
M$_{\odot}$ yr$^{-1}$ pc$^{-2}$ and for the southern spiral arm is
$0.23\times10^{-8}$ M$_{\odot}$ yr$^{-1}$ pc$^{-2}$.   This suggests that the
parts of the disk of NGC~2805 from which radio continuum emission at 1.4 GHz is
detected show uniform rate of star formation throughout the disk.  Normal
star forming galaxies show SFR which range from $7.4\times10^{-10}$ to
$1.6\times10^{-7}$  M$_{\odot}$ yr$^{-1}$ pc$^{-2}$ \citep{kennicutt98} and NGC~2805 
falls in this range as expected.

\begin{table}
\begin{center}
\caption{Properties of the two large spiral galaxies in the group.}
\begin{tabular}{@{}lll@{}}\hline
Parameters& NGC~2805&NGC~2820\\  \hline
H{\sc i} flux density (Jy~km s$^{-1}$)&86.9$\pm$1.2& 49.4$\pm$1.1\\
H{\sc i} mass (10$^{9}$ $M_{\odot}$)&12.5 &  7.1\\
Flux density at $_{1.4}$ GHz (mJy) & 19.5$\pm$0.3$^{a}$&90$\pm$0.26$^{b}$ \\
FIR to radio correlation (q)&2.2$\pm$0.02&1.8$\pm$0.03\\
SFR$_{1.4MHz}$$^{c}$ ($M_{\odot}$/yr)&0.86$\pm$0.5&3.9$\pm$0.2 \\
Integrated flux density at 325~MHz(mJy)&73.1$\pm$0.9 & 227 \\
Spectral Index ($\alpha^{325}_{1420}$) & -0.8$\pm$0.2& -0.6$^{d}$ \\   \hline
\end{tabular}\\[5pt]
\begin{minipage}{10 cm}
\small Notes: The result at 325 MHz are taken from \citep{kantharia05}.
             (a) The flux density of NGC~2805 is estimated from the 
                 NVSS 1.4 GHz image.   
             (b) Flux density at 1.4 GHz has been estimated for 
                 NGC~2820+Mrk~108 using our 1.4 GHz radio continuum map.
             (c) SFR$_{1.4GHz}$ drive from the  \citep{yun01}.
             (d) Since its difficult to separate the flux density of 
                 NGC~2820 and Mrk~108, The flux density and spectral 
                 index includes contribution from both the galaxies.

\end{minipage}
\end{center}
\end{table}

\subsection{The model}
Kantharia et. al (2005) have proposed a model for Holmberg~124; especially
the triplet galaxies namely NGC~2820, NGC~2814 and Mrk~108; 
according to which both tidal and ram pressure effects are operative
in this triplet galaxies and also likely for NGC~2805. 
Our analysis of the radio continuum
and the 21cm H{\sc i} images support their scenario for the triplet.   The tidal features that
we detect are (i) the radio continuum and H{\sc i} bridge connecting NGC
2820, Mrk~108 with NGC~2814, (ii) the streamer spatially connected to
NGC~2814 but kinematically distinct from it and (iii) the tidal dwarf
galaxy to the north-east of NGC~2820. Further, as discussed below, our
analysis supports their view regarding the
triplet viz., NGC~2820 has dominant motion in the south-east and NGC
2814 to the north in the sky plane. This motion in the IGrM,
with a small line-of-sight component, leads to
ram pressure effects leaving detectable morphological imprints on the 
member galaxies in the atomic H{\sc i} and radio continuum emission in the sky
plane.

We estimate an H{\sc i} deficiency of $-0.10$ for NGC~2805 and $-0.04$ for NGC~2820,
indicating that the two giant spirals are not H{\sc i}-deficient and hence it is likely
they have just started experiencing ram pressure effects which are observed
as morphological signatures.  For estimating the deficiency, we have used the
prescription in which the H{\sc i} content of a particular galaxy type is compared to
the H{\sc i} content in a field galaxy as given by Haynes \& Giovanelli (1984).  
For NGC~2805, we used type 4 and for NGC~2820 we
used the H{\sc i} content of type 5 galaxy.   
Since we have been able to obtain better H{\sc i} images of the low inclination
galaxy NGC~2805, we can comment on the physical processes 
which the galaxy is likely experiencing and which in turn can explain
the observed H{\sc i} distribution.  We can explain both the sharp fall in the 
H{\sc i} column density
in the north and the star forming ridge along the spiral arm in the south
due to the motion of the galaxy in the IGrM.
Enhanced star formation is observed in the nuclear regions 
of the galaxy \citep{ganda06}.  Moreover while the NIR emission is predominantly
centrally concentrated; the UV and H$\alpha$ emission 
arise from several regions in the galaxy and is especially intense in the
southern spiral arm.  Based on all this, we suggest that NGC~2805 is moving 
along a direction at a small angle to the
line-of-sight and tilted towards the triplet i.e. north-west.  The ram pressure effects
on NGC~2805 is due to an almost face-on encounter with the IGrM with the northern
H{\sc i} arc being the leading edge.  This is similar to 
the model given for NGC 4254 (\citealt{vollmer05}; \citealt{kantharia08}) where almost
the entire disk of the galaxy is encountering the IGrM 
in a face-on event which leads to compression and
enhanced star formation in the southern arm.
The star formation in the southern spiral arm in NGC~2805 has been triggered
by the large gas densities encountering the IGrM.  
The motion of the galaxy and the subsequent ram pressure effects 
leads to gas being stripped out of the disk
which explains the tenuous H{\sc i} gas detected in the south-west. 
Simulations e.g. \citet{roediger06} where the galaxy has an face-on encounter
with the IGrM also show the gas being stripped off and appearing to form a gas halo
around the galaxy in the sky plane. 
As suggested by Kantharia et al. (2005), the motion of NGC~2820 and
NGC~2814 is predominantly in the sky plane with a small line-of-sight component.
This is supported by the systemic velocity of all
the components.  The velocities of NGC~2820 and 
NGC~2814 are 1574 and 1592 km~s$^{-1}$ respectively   
while the velocity of NGC~2805 is 1725 km~s$^{-1}$ - this 
supports our model that its motion is predominantly radial and receding.

\section{Summary}\label{s:consl}
In this paper, we have presented new H{\sc i} images of the member galaxies of the
group Holmberg~124 made from archival GMRT 21cm H{\sc i}  data. These images confirm
the results of Kantharia et al. (2005) for the triplet consisting of
the large spiral NGC~2820 and the smaller late type galaxies NGC~2814 and 
Mrk~108 including the detection of a possible tidal dwarf galaxy and the large
asymmetric H{\sc i} loop seen along the minor axis in NGC~2820.  Moreover,  superior 
quality H{\sc i} emission images of NGC~2805 are presented and comparison with single dish 
data shows that the GMRT data has recovered most of the H{\sc i} mass for this galaxy with 
large angular size with the GMRT estimate being lower by only about 5\%.  
NGC~2805 is a low inclination galaxy and its optical emission is characterized by 
a star forming ridge along a spiral arm in the south and a star forming arc in the 
northern part in addition to star formation throughout the galaxy.  Our H{\sc i} column density 
maps show enhanced column densities in the northern arc and in the southern 
spiral arm with H{\sc i} from the entire optical disk and diffuse H{\sc i} detected 
beyond the southern arm. This diffuse H{\sc i} shows peculiar morphology and has low 
velocity dispersion compared to rest of the galaxy.  It also seems to be 
losing rotation. Based on these properties, we suggest that this gas
might not be co-planar to the rest of the disk emission but might be being stripped
off the galaxy as it moves through the IGrM. 
From the combined nature of the H{\sc i} and optical morphologies, we suggest that NGC~2805 
is showing clear signatures of ram pressure effects resulting from it motion along a 
direction close to the line-of-sight but tilted towards the triplet i.e. towards north-east.  
This is supported by the systemic velocities of the galaxies - the velocities of the 
triplet galaxies are similar whereas that of NGC~2805 differs from these by more than 
100 km~s$^{-1}$. So while the triplet have dominant motion in the sky plane, NGC~2805 
has dominant motion along the line-of-sight direction.   The compressed H{\sc i} in the north, 
the star formation triggered throughout the galaxy and especially the southern spiral arm 
and the diffuse gas to the south of the galaxy are all explained in this model, 
since the entire disk of NGC~2805 would  encounter the IGrM.  
Neither of the large spiral galaxies are H{\sc i} deficient indicating 
that they are probably in the initial stages of experiencing ram pressure effects.
We note that this is an interesting group of late type galaxies with one galaxy
seen at high inclination and the other at low inclination allowing us to
study a range of features due to ram pressure effects. 
We also report tentative detection of H{\sc i} in the IGrM in the form of small 
discrete clumps near NGC~2820.   The clumps near 
NGC~2820 are detected near the asymmetric loop and south of the triplet. 
We note that these features need confirmation
since their detection in the GMRT data is marginal.  It will be very interesting
to follow up these interesting results with deep X-ray observations of
the IGrM and H{\sc i} observations sensitive to larger angular scales.   It is necessary to observe many 
more such groups to understand the physical mechanisms which control the evolution 
of the member galaxies, especially since more than 60\% of all galaxies
are believed to be in group environs.  While tidal interactions are believed to be active in such 
systems; there are increasing number of systems where ram pressure effects seem
to be showing observable effects. 

\section*{Acknowledgements}
We thank the staff of the GMRT who made this observations possible. The GMRT is operated by 
the National Centre for Radio Astrophysics of the Tata Institute of Fundamental Research.  
A. Mishra thanks National Centre for Radio Astrophysics for 
support and hospitality.  We thank S. Ananthakrishnan for suggesting that we examine
the archival data on this group. We would also like to thank D.J. Saikia, Editor, BASI and 
an anonymous referee for giving several
useful suggestions. This research has made use of NASA/IPAC Infrared Science Archive 
and the NASA/IPAC Extragalactic Data base (NED) both of which are operated by 
Jet Propulsion Laboratory, California Institute of Technology under contract 
with the National Aeronautics and Space Administration.


\begin{thebibliography}{}
\bibitem[\protect\citeauthoryear{Artamonov}{1994}]{artamonov94} Artamonov B. P., 1994, IAUS, 161, 592
\bibitem[\protect\citeauthoryear{Boker et al.}{2002}]{boker02} Boker T., Laine S., van der M., 
Roeland P., Sarzi M., Rix H. W., Ho L. C., Shields J. C., 2002,  AJ, 123, 1389B
\bibitem[\protect\citeauthoryear{Bosma et al.}{1980}]{bosma80} Bosma A., Casini C., Heidmann J., 
van der Hulst J. M., van Woerden  H., 1980, A\&A, 89, 345
\bibitem[\protect\citeauthoryear{Chung et al.}{2007}]{chung07} Chung A., van Gorkom J. H.,
Kenney J. D. P., Vollmer B., 2007, ApJ, 659, 115
\bibitem[\protect\citeauthoryear{Dahlem}{2005}]{dahlem05} Dahlem M., 2005, A\&A, 429, 5
\bibitem[\protect\citeauthoryear{Fitt et al.}{1993}]{fitt93} Fitt J. A., Alexander P., 1993, MNRAS, 261, 445
\bibitem[\protect\citeauthoryear{Ganda et al.}{2006}]{ganda06} Ganda K., Falcon-Barroso  J., Peletier 
R. F., Cappellari M., Emsellem E., McDermid R. M., de Zeeuw P. T., Carollo C. M., 2006, MNRAS, 367, 46G
\bibitem[\protect\citeauthoryear{Garrido et al.}{2004}]{garrido04} Garrido O., Marcelin M., Amram P., 2004, 
MNRAS, 349, 225
\bibitem[\protect\citeauthoryear{Gunn \& Gott}{1972}]{gunn72} Gunn J. E., Gott III J. R., 1972,
ApJ, 176, 1
\bibitem[\protect\citeauthoryear{Haynes \& Giovanelli}{1984}]{haynes84} Haynes M. P., Giovanelli R., 1984, AJ, 
89, 758
\bibitem[\protect\citeauthoryear{Hodge}{1975}]{hodge75} Hodge W. P., 1975, ApJ, 201, 556
\bibitem[\protect\citeauthoryear{van der Hulst et al.}{1985}]{van der hulst85} van der Hulst J. M., 
Hummel E.,  1985, A\&A, 150L, 7V
\bibitem[\protect\citeauthoryear{Kantharia et al.}{2005}]{kantharia05} Kantharia N. G., Ananthakrishnan S.,
Nityananda R., Hota A., 2005, A\&A, 435, 483
\bibitem[\protect\citeauthoryear{Kantharia et al.}{2008}]{kantharia08} Kantharia N. G., Rao A. Pramesh,
Sirothia S. K., 2008, MNRAS, 383, 173
\bibitem[\protect\citeauthoryear{Kennicutt}{1998}]{kennicutt98} Kennicutt Robert C. Jr., 1998,
ApJ, 498, 541
\bibitem[\protect\citeauthoryear{Kilborn et al.}{2006}]{kilborn06} Kilborn V. A., Forbes D. A.,
Koribalski B. S., Brough S., Kern K., 2006, MNRAS, 371, 739
\bibitem[\protect\citeauthoryear{Lang et al.}{2003}]{lang03} Lang H. R., Boyce P. J., Kilborn V. A.,
Minchin R. F., Disney M. J., Jordan C. A., Grossi M., Garcia D. A., Freeman K. C., Philipps S.,
Wright A. E., 2003,  MNRAS, 342, 738L
\bibitem[\protect\citeauthoryear{Moore et al.}{1996}]{moore96} Moore B., Katz N., Lake G.,
Dressler A., Oemler A.,  1996, Natur., 379, 613
\bibitem[\protect\citeauthoryear{Nulsen}{1982}]{nulsen82} Nulsen P. E. J., 1982, MNRAS, 198, 1007
\bibitem[\protect\citeauthoryear{Rasmussen et al.}{2012}]{rasmussen12} Rasmussen J., Bai X. N.,
Mulchaey J. S., van Gorkom J. H., Jeltema T. E., Zabludoff A. I., Wilcots E., Martini P.,
Lee D., Roberts T. P., 2012, ApJ, 747, 31
\bibitem[\protect\citeauthoryear{Reakes}{1979}]{reakes79} Reakes M., 1979,  MNRAS, 187, 525
\bibitem[\protect\citeauthoryear{Roediger \& Bruggen}{2006}]{roediger06} Roediger E., Bruggen 
M.,  2006,  MNRAS, 369, 567
\bibitem[\protect\citeauthoryear{Sengupta et al.}{2007}]{sengupta07} Sengupta C., 
Balasubramanyam R., Dwarakanath K. S., 2007,  MNRAS, 378, 137
\bibitem[\protect\citeauthoryear{Swarup et al.}{1991}]{swarup91} Swarup G.,
Ananthakrishnan S., Kapahi V. K., Rao A. P., Subrahmanya C. R., Kulkarni V. K., 1991, 
Current Science, 60, 95
\bibitem[\protect\citeauthoryear{Toomre \& Toomre}{1972}]{toomre72} Toomre A., Toomre J., 1972, 
ApJ, 178, 623
\bibitem[\protect\citeauthoryear{Toomre}{1977}]{toomre77} Toomre  A., 1977, in Beatrice M.,
Tinsley and Richard B. Larson, eds. Evolution  of Galaxies and Stellar Population, 
New Haven, Yale University Observatory, p. 401
\bibitem[\protect\citeauthoryear{Verdes-Montenegro et al.}{2001}]{verdes01} Verdes-Montenegro L.,
Yun M.S., Williams B.A., Huchtmeier W. K., Del Olmo A., Perea J., 2001, A\&A, 377, 812
\bibitem[\protect\citeauthoryear{Vollmer et al.}{2001}]{vollmer01} Vollmer B., Cayatte V.,
Balkowski C., Duschl W. J., 2001, ApJ, 561, 708
\bibitem[\protect\citeauthoryear{Vollmer et al.}{2005}]{vollmer05} Vollmer B., Huchtmeier
W., van Driel W.,  2005, A\&A, 439, 921
\bibitem[\protect\citeauthoryear{Yun et al.}{2001}]{yun01} Yun M.S., Reddy N.A., Condon J.J.,  2001, 
ApJ, 554, 803
\end{thebibliography}
\end{document}